\documentclass[preprintnumbers,amsmath,amssymb,showpacs]{revtex4}
\usepackage{epsfig}
\usepackage{color}
\usepackage[colorlinks,urlcolor=blue,citecolor=blue]{hyperref}
\def\hhref#1{\href{http://arxiv.org/abs/#1}{arXiv:#1}} 
\pdfoutput=0
\begin{document}
\setlength{\voffset}{1.0cm}
\title{Transparent Dirac potentials in one dimension: the time-dependent case}
\author{Gerald V. Dunne}
\affiliation{ARC Centre of Excellence in Particle Physics at the Terascale and CSSM,
School of Chemistry and Physics, University of Adelaide, Adelaide SA 5005, Australia;
\\Physics Department, University of Connecticut, Storrs CT 06269, USA}
\author{Michael Thies}\affiliation{Institut f\"ur  Theoretische Physik, Universit\"at Erlangen-N\"urnberg, D-91058,Erlangen, Germany}
\date{\today}
\begin{abstract}We generalize the original derivation of transparent, static Schr\"odinger potentials by Kay and Moses, to obtain a large 
class of {\it time-dependent} transparent Dirac potentials in one spatial dimension. They contain all known transparent potentials as special 
cases and play a key role in the semi-classical solution of 1+1 dimensional, fermionic quantum field theories of Gross-Neveu and 
Nambu--Jona-Lasinio type.
\end{abstract}
\maketitle
\section{Introduction}
\label{sect1}
Transparent, static potentials of the one-dimensional Schr\"odinger equation were first characterized systematically in the 
seminal paper by Kay and Moses (KM) \cite{1}. These are potentials with the property that the reflection coefficient for a scattered particle 
vanishes at all energies. KM  constructed the most general potential of this type by algebraic methods \cite{1}. Their proof that this was 
indeed the most general transparent potential was based on inverse scattering theory and the Gel'fand-Levitan-Marchenko equation
(for a review see \cite{1a}). One can also define transparent, time dependent potentials. The reason is the fact that the asymptotic behavior 
of the scattering wavefunctions is the same as for static potentials, if one takes the limit $x\to \pm \infty$ at a fixed time. The generalization 
of the KM work to time dependent Schr\"odinger potentials can be found in work of Nogami and Warke on soliton solutions of the 
multicomponent nonlinear Schr\"odinger equation \cite{2}. Although these authors did not address the issue of transparency,  it  is easy to 
find the scattering solutions to their time dependent potential and confirm that these are reflectionless. As a matter of fact, the continuum 
wave functions  (in a notation adapted to that of \cite{2}) are given by
\begin{equation}
g_k = \left( 1 + \sum_{\alpha} \frac{u_{\alpha}^* g_{\alpha}}{ik+\kappa_{\alpha}^*} \right) e^{i(kx-k^2t)}
\label{1}
\end{equation}
where the $g_{\alpha}$ are bound state wave functions, and the $u_{\alpha}$ are exponentials in $x,t$.

The KM results have also been generalized to the relativistic regime. Toyama, Nogami and Zhao \cite{3} solve a stationary, nonlinear
Dirac equation by reducing it to a pair of (supersymmetric) Schr\"odinger equations. This enables them to derive all static, 
Lorentz-scalar, transparent Dirac potentials from the known KM potentials. Their results agree with independent work on the 
large-$N$ solution of baryons and multibaryon bound states in the Gross-Neveu (GN) model \cite{4,5,6}. In this context, it has long been
known that self-consistent mean field potentials are necessarily transparent \cite{5}. The authors of \cite{3} argue heuristically that
Lorentz-vector potentials are unlikely to be transparent. They also point out that their method fails in the non-static case, simply 
because one cannot reduce the time-dependent  Dirac equation to a Schr\"odinger equation. Later on, Nogami and Toyama reported
on transparent, pseudoscalar Dirac potentials as well \cite{7}. With hindsight, their search was too narrow, being restricted to an
$x$-dependent pseudoscalar and a constant scalar potential. This situation only changed very recently with a series of works by
Nitta, Takahashi and coworkers \cite{8,9,10}. Working on self-consistent, static solutions of the Nambu--Jona Lasinio (NJL$_2$) model
or, equivalently, the Bogoliubov-de Gennes (BdG) equation, they found a very general class of static, transparent scalar-pseudoscalar
potentials of the one-dimensional Dirac equation. These correspond to bound states of any number of chirally twisted kinks of the
type first discussed by Shei in the context of the NJL$_2$ model \cite{11}. 

If it is indeed correct that only scalar and pseudoscalar Dirac potentials have a chance of being transparent, the only step missing so 
far is the generalization of these relativistic transparent potentials to the time dependent case. 
This problem is intimately related to the self-consistent solution of dynamical problems in the GN and NJL$_2$ models, such as 
scattering of kinks, breathers and bound states thereof. Lorentz scalar potentials are relevant for time-dependent Hartree-Fock (TDHF) 
solutions of the large-$N$ GN model. Scalar-pseudoscalar potentials are needed to solve the NJL$_2$ model or the BdG equation in 
condensed matter physics, where chiral twists are possible. It is the purpose of this paper to fill this gap and construct a general class 
of transparent scalar and pseudoscalar, time dependent Dirac potentials. Related results have already been presented in 
a recent letter \cite{12}, and here we give an even simpler characterization of time-dependent transparent Dirac potentials. The detailed 
application of the results obtained here to exactly solvable quantum field theories will be the subject of a forthcoming publication \cite{13}.
\section{Formalism and Results}
\label{sect2}
\subsection{Formalism}
Our starting point is the Dirac equation in one space dimension with real scalar ($S$) and pseudoscalar ($P$)
potentials depending on $x$ and $t$,
\begin{equation}
(i \partial \!\!\!/ - S -i \gamma_5 P)\psi = 0.
\label{2}
\end{equation}
We choose a chiral representation of the Dirac matrices,
\begin{equation}
\gamma^0 = \sigma_1, \quad \gamma^1 = i \sigma_2, \quad \gamma_5 = \gamma^0 
\gamma^1 = - \sigma_3,
\label{3}
\end{equation}
and go over to light cone coordinates,
\begin{equation}
z=x-t, \quad \bar{z} = x+t, \quad
\partial_0 = \bar{\partial}-\partial, \quad \partial_1 = \bar{\partial} + \partial .
\label{4}
\end{equation}
(Note that  $\bar{z}$ is {\it not} the complex conjugate of $z$, which we would write as $z^*$.) 
Momentum $k$ and energy $E$ are encoded in the light cone spectral parameter (using natural units where the fermion mass is set to 1),
\begin{equation}
k = \frac{1}{2}\left( \zeta- \frac{1}{\zeta} \right), \quad E = - \frac{1}{2} \left( \zeta + \frac{1}{\zeta} \right)
\label{5}
\end{equation}
In these variables, the Lorentz scalar argument of a plane wave reads
\begin{equation}
k_{\mu} x^{\mu} = - \frac{1}{2} \left( \zeta \bar{z}- \frac{z}{\zeta} \right).
\label{6}
\end{equation}
The Dirac equation in light cone coordinates can be written in components as 
\begin{equation}
2i \bar{\partial}\psi_2 = \Delta \psi_1, \quad 2i \partial \psi_1 = -\Delta^* \psi_2, 
\label{7}
\end{equation}
where the scalar and pseudoscalar potentials are combined into a single complex potential 
$\Delta$:
\begin{equation}
\Delta \equiv S-iP.
\label{7b}
\end{equation}
In Eq (\ref{7}), $\psi_1, \psi_2$ are the upper and lower spinor components with left- and right-handed chirality, respectively.
\subsection{Ansatz}
In order to find transparent potentials $\Delta$, we start from the following ansatz for the continuum spinor,
\begin{equation}
\psi_{\zeta} = \frac{1}{\sqrt{1+\zeta^2}} \left( \begin{array}{c} \zeta \chi_1 \\ - \chi_2 \end{array} \right) e^{i(\zeta \bar{z}-z/\zeta)/2}.
\label{8}
\end{equation}
This ansatz has a standard continuum normalization for free, massive spinors provided that $\lim_{x\to -\infty} \chi_{1,2} = 1$.
We also demand that $\chi_1$ and $\chi_2$ approach some constant for $x \to \infty$. In that case, the continuum 
spinor behaves like a plane wave travelling to the right for $x\to -\infty$ as well as for $x\to \infty$ (for $k>0$), hence it is manifestly
reflectionless. The Dirac equation for the reduced spinor components $\chi_1,\chi_2$ takes on the form
\begin{eqnarray}
(2 i \bar{\partial}-\zeta)\chi_2 + \zeta \Delta \chi_1 & = & 0 ,
\label{9} \\
( 2i \zeta  \partial + 1) \chi_1 - \Delta^* \chi_2 & = & 0.
\label{10}
\end{eqnarray}
We introduce $N$ basis functions $e_n, f_n$, reminiscent of plane waves, but with complex spectral parameters,
\begin{equation}
e_n = e^{i(\zeta_n^* \bar{z} - z/\zeta_n^*)/2}, \quad f_n = \frac{e_n}{\zeta_n^*}, \quad \zeta_n \in {\mathbb C}, \quad n=1,...,N
\label{11}
\end{equation}
As our ansatz to construct a transparent potential we assume that $\chi_{1,2}$ can be represented 
as a finite sum  of these basis functions, with a finite number, $N$, of poles in the complex spectral plane:
\begin{eqnarray}
\chi_1 & = &   
 1 + i \sum_{n=1}^N \frac{1}{\zeta-\zeta_n} e_n^* \phi_{1,n} ,
\nonumber \\
\chi_2 & = &    1 - i \sum_{n=1}^N \frac{\zeta}{\zeta-\zeta_n} e_n^* \phi_{2,n}, 
\label{12}
\end{eqnarray}
where $\phi_{1,n}$ and $\phi_{2,n}$ are 2$N$ functions defined as the solutions of the following systems of {\it linear, algebraic} 
equations,
\begin{eqnarray}
\sum_{m=1}^N \left( \omega+B\right)_{nm}\phi_{1,m} & = &  e_n,
\nonumber \\
\sum_{m=1}^N \left( \omega+B\right)_{nm}\phi_{2,m} & = & - f_n.
\label{13}
\end{eqnarray}
Here, $\omega$ is a constant, hermitean but otherwise arbitrary $N\times N$ matrix, and $B$ is an $N\times N$ matrix constructed from the basis functions, 
$e_n(z, \bar z)$, and spectral parameters, $\zeta_n$, as follows:
\begin{equation}
B_{nm} = i \frac{e_n e_m^*}{\zeta_m-\zeta_n^*}.
\label{14}
\end{equation}
From Eq.~(\ref{12}) we see that the parameters $\zeta_n$ introduced in (\ref{11})  correspond to the positions of the bound state 
poles of $\psi_{\zeta}$ in the complex $\zeta$-plane.
\subsection{Proof of Solution}
To show that the ansatz above
provides an exact time-dependent solution of the Dirac equation, with a transparent potential $\Delta$, we use the following 
elementary algebraic steps. We denote by $e, f, \phi_1, \phi_2$ the $N$-dimensional column vectors with components 
$e_n,f_n,\phi_{1,n},\phi_{2,n}$, respectively, whereas $\omega$ and $B$ denote $N \times N$ matrices. Eq.~(\ref{13}) becomes 
the pair of matrix equations:
\begin{eqnarray}
 ( \omega + B ) \phi_1 & =  & e,
 \label{15} \\ 
( \omega + B ) \phi_2 & = & - f.
 \label{16}
\end{eqnarray}
Differentiating the column vectors $e$ and $f$, using the definitions (\ref{11}), leads to the simple relations
\begin{equation}
2 i \partial e = f, \quad 2i \bar{\partial} f = - e,
\label{17}
\end{equation}
showing that the $N$ spinors with components $(e_n,-f_n)^T$ are (unnormalizable, due to the complex spectral parameter) solutions
of the free, massive Dirac equation ($m=1$). This then implies an important property of the matrix $B$: its
derivatives with respect to $z,\bar{z}$ are separable matrices,
\begin{equation}
\partial B = \frac{1}{2} f f^{\dagger}, \quad \bar{\partial} B = \frac{1}{2} e e^{\dagger}.
\label{18}
\end{equation}
Let us apply $2i \partial$ to (\ref{15}), and $2i \bar{\partial}$ to (\ref{16}). Using (\ref{17}, \ref{18}), we find
\begin{eqnarray}
(\omega+ B) 2i \partial \phi_1 & = & f (1-i f^{\dagger} \phi_1) 
,\nonumber \\
(\omega+ B) 2i \bar{\partial} \phi_2 & = & e (1 - i e^{\dagger} \phi_2) .
\label{19}
\end{eqnarray}
Upon multiplying these equations from the left by $(\omega+ B)^{-1}$, we get
\begin{equation}
2 i \partial \phi_1 = - \phi_2 (1-i f^{\dagger}\phi_1), \quad 2 i \bar{\partial} \phi_2 = \phi_1 (1- i e^{\dagger} \phi_2),
\label{20}
\end{equation}
showing that the $\phi_{1,n}, \phi_{2,n}$ are the upper and lower components of $N$ distinct solutions of the Dirac equation with 
potential
\begin{equation}\Delta  =   1 - i e^{\dagger} \phi_2 = 1 + i \phi_1^{\dagger} f = 1+i e^{\dagger} \frac{1}{\omega+B} f .
\label{21}
\end{equation} 
The three different expressions for $\Delta$ given here in (\ref{21}) are equivalent owing to Eqs.~(\ref{15}, \ref{16}).
The $\phi_1,\phi_2$ are (normalizable) bound state spinors of the potential $\Delta$. If we insert them into our ansatz
for the continuum spinor (\ref{12}), it is now straightforward to verify that the continuum spinor satisfies the Dirac equation 
(\ref{9}, \ref{10}) with the potential (\ref{21}), proving that this potential is transparent. In view of the crucial importance of this
last step, we give a few details of the calculation. Let us spell out the three terms appearing in Eq.~(\ref{9}),
\begin{eqnarray}
2i \bar{\partial} \chi_2 & = & - i \sum_{n} \frac{\zeta}{\zeta-\zeta_n} e_n^* \left( \zeta_n \phi_{2,n} + \Delta \phi_{1,n} \right),
\nonumber \\
- \zeta \chi_2 & = & - \zeta + i \sum_n \frac{\zeta^2}{\zeta-\zeta_n} e_n^* \phi_{2,n},
\nonumber \\
\zeta \Delta \chi_1 & = & \zeta \Delta + \zeta \Delta i \sum_n \frac{1}{\zeta-\zeta_n} e_n^* \phi_{1,n},
\label{22}
\end{eqnarray}
where we have carried out $\bar{\partial} e_n^*$, and used the Dirac equation to evaluate $\bar{\partial}\phi_{2,n}$.
To obtain a solution, these three terms should add up to zero. The terms containing $\phi_{1,n}$ in the 1st and 3rd lines 
of (\ref{22}) clearly cancel. The terms containing $\phi_{2,n}$ in the 1st and 2nd lines of (\ref{22}) add up to $\zeta(1-\Delta)$, 
using Eq. (\ref{21}), and therefore cancel the remaining terms in the 2nd and 3rd lines of (\ref{22}). Likewise, the three terms 
of Eq.~(\ref{10}) become
\begin{eqnarray}
2i \zeta \partial \chi_1 & = & - i \sum_{n} \frac{\zeta}{\zeta-\zeta_n} \left( f_n^* \phi_{1,n} + e_n^* \Delta^* \phi_{2,n} \right)
\nonumber \\
\chi_1 & = & 1 + i \sum_n \frac{1}{\zeta-\zeta_n} e_n^* \phi_{1,n}\nonumber \\-  \Delta^* \chi_2 & = & - \Delta^* + i \Delta^*  
\sum_n \frac{\zeta}{\zeta-\zeta_n} e_n^* \phi_{2,n}
\label{23}
\end{eqnarray}
Here, the terms containing $\phi_{2,n}$ in the 1st and 3rd lines cancel. The terms containing $\phi_{1,n}$ in the 1st and 2nd lines 
add up to $(\Delta^*-1)$, and cancel the remaining terms in the 2nd and 3rd lines.

This completes the proof that we have indeed 
found a whole class of transparent, time-dependent potentials of the Dirac equation. The potential $\Delta$ is given 
by (\ref{21}) in terms of the basis functions $e_n$ and/or $f_n$,  and the solutions of the algebraic equations (\ref{15}, \ref{16}). These 
time-dependent transparent potentials $\Delta(x, t)$ are parameterized by $N$ complex constants $\zeta_n$ related to $N$ bound 
state poles in the complex $\zeta$-plane, and by a constant, hermitean $N \times N$ matrix $\omega$. Further conditions 
can be placed on the matrix $\omega$ to ensure that the potential $\Delta$ is physical; for example, to ensure that the solution does 
not develop singularities as a function of $x$ and $t$, or violate cluster separability. Details of choosing physical scattering solutions
will be discussed in a separate work \cite{13}. 

We have not yet been able to prove that our result is the most general transparent potential. 
The tools used by KM in their corresponding proof for the static Schr\"odinger potentials, namely inverse scattering theory and the 
Gel'fand-Levitan-Marchenko equation, are not available in the present time-dependent case.
\section{General properties of the result}
\label{sect3}
The transparent potentials derived above in Eq. (\ref{21}) describe a rich variety of solitons,
multi-soliton bound and scattering states as well as states involving breathers of increasing complexity. Since these solitons play the 
role as TDHF potentials in integrable quantum field theories like the GN and the NJL$_2$ models, we
postpone a detailed discussion to a forthcoming paper, where also the self-consistency will be shown explicitly \cite{13}. Rather than 
studying concrete examples, we collect in this section some general properties of our solution which will turn out to be helpful in analyzing 
the above mentioned integrable field theoretic models.
\subsection{Determinants}
\label{sect3a}
Since the bound state spinors can be regarded as residues of the continuum spinor at the bound state poles $\zeta_n$, essentially all
 information is contained in the three quantities $\Delta, \chi_1, \chi_2$:  recall that $\Delta(x, t)$ is the complex potential, whose
 real and imaginary parts give the scalar and pseudoscalar Dirac potentials, respectively, while $\chi_{1, 2}$ define the chiral components
 of the associated continuum spinors in (\ref{8}). Our algebraic construction leads to simple and explicit representations of these key 
quantities in terms of determinants. Let us define the following $N \times N$ matrices, related to the matrix $B$ defined in Eq.~(\ref{14}):
\begin{eqnarray}
A_{nm} & = & \frac{\zeta_m}{\zeta_n^*} B_{nm},
\nonumber \\
C_{nm} & = & \frac{\zeta-\zeta_n^*}{\zeta-\zeta_m} B_{nm} ,
\nonumber \\
D_{nm} & = & \frac{\zeta_m}{\zeta_n^*} C_{nm},
\label{24}
\end{eqnarray}
The distinguishing feature of these three matrices is the fact that they differ from $B$ only through separable matrices,
\begin{eqnarray}
A-B & = & i f e^{\dagger},
\nonumber \\
C-B & = & i e g^{\dagger}, 
\nonumber \\
D-B & = & i \zeta f g^{\dagger} .
\label{25}
\end{eqnarray}
Here we have introduced a 3rd vector $g$, in addition to the vectors $e, f$ defined in Eq.~(\ref{11}), with components 
 \begin{equation}
g_n = \frac{e_n}{\zeta-\zeta_n^*}.
\label{26}
\end{equation}
The determinant of the sum of an invertible matrix $M$ and a separable matrix $a\, b^{\dagger}$ can be computed as follows \cite{13a}:
\begin{equation}
\frac{{\rm det} (M+ a\, b^{\dagger})}{{\rm det} (M)} = 1 + b^{\dagger} \frac{1}{M} a .
\label{27}
\end{equation}
Applying this algebraic identity to the quantities $\Delta, \chi_1, \chi_2$, using their expressions in (\ref{21}, \ref{12}), yields
\begin{eqnarray}
\Delta & = &  1 + i e^{\dagger}\frac{1}{\omega+B} f = \frac{{\rm det}(\omega+A)}{{\rm det}(\omega + B)}   , 
\nonumber \\
\chi_1  & = &    1 + i g^{\dagger} \frac{1}{\omega+B} e = \frac{{\rm det}(\omega+C)}{{\rm det}(\omega + B)}  ,
\nonumber \\
\chi_2  & = &   1 + i \zeta g^{\dagger} \frac{1}{\omega+B} f = \frac{{\rm det}(\omega+D)}{{\rm det}(\omega + B)}  .
\label{28}
\end{eqnarray}
Eq. (\ref{28}) is a compact expression of our main result: we have expressed 
each of the three key quantities, the potential and the components of the continuum spinors, as a ratio of determinants of 
simple matrices. 
\subsection{Diagonal matrix $\omega$: scattering of chirally  twisted kinks}
\label{sect3b}
The expressions (\ref{28}) can be further simplified in the case where the matrix $\omega$ is diagonal. As will be discussed in more detail 
elsewhere \cite{13}, this corresponds to the situation where $N$ twisted kinks scatter without forming breathers. In this special
case,  because of the extra symmetry of the mixing matrix $\omega$, it is advantageous to introduce $N$ functions $U_n$, and
a $N \times N$ coefficient matrix $b_{nm}$, as follows,
\begin{equation}
U_n = \frac{i |e_n|^2}{\omega_{nn}(\zeta_n-\zeta_n^*)}, \qquad b_{nm}= \left| \frac{\zeta_n-\zeta_m}{\zeta_n - \zeta_m^*}\right|^2.
\label{28a}
\end{equation}
The determinant of $\omega+B$ can then be reduced to the following simple expression,
\begin{equation}
\frac{{\rm det} (\omega+B)}{{\rm det} \omega} = 1 + \sum_n U_n + \sum_{n<m} b_{nm} U_n U_m + \sum_{n<m<k} 
b_{nm}b_{nk}b_{mk} U_n U_m U_k + ... +\prod_{n<m} b_{nm} \prod_k U_k.
\label{28b}
\end{equation}  
Furthermore, the other determinants appearing in (\ref{28}) can be inferred from this expression by merely rescaling
the variables $U_n$ by complex factors,
\begin{eqnarray}
{\rm det} (\omega+A) & = & {\rm det} (\omega+B) \left( U_n \to \frac{\zeta_n}{\zeta_n^*} U_n \right) ,
\nonumber \\
{\rm det} (\omega+C) & = & {\rm det} (\omega+B) \left( U_n \to \frac{\zeta-\zeta_n^*}{\zeta-\zeta_n} U_n \right) ,
\nonumber \\
{\rm det} (\omega+D) & = & {\rm det} (\omega+B) \left( U_n \to   \frac{\zeta_n}{\zeta_n^*}  \frac{\zeta-\zeta_n^*}{\zeta-\zeta_n} U_n \right) .
\label{28c}
\end{eqnarray}
If we further restrict this special case to the situation where the complex spectral parameters $\zeta_n$ all lie on the unit 
circle, these determinant expressions (\ref{28c}) provide an explicit closed-form solution to the finite algebraic system of equations 
used in \cite{8, 9, 10} to define {\it static} transparent  Dirac potentials.
\subsection{Diagonal matrix $\omega$ and pure imaginary spectral parameters $\zeta_n$: scattering of real  kinks}
\label{sect3c}
A further simplification is obtained by choosing the mixing matrix $\omega$ to be diagonal, as in the previous subsection, 
but restricting the spectral parameters to be purely imaginary. In this case, the determinant expressions in (\ref{28}) reduce to:
\begin{eqnarray}
{\rm det} (\omega+A) & = & {\rm det} (\omega+B) \left( U_n \to -  U_n \right) ,
\nonumber \\
{\rm det} (\omega+C) & = & {\rm det} (\omega+B) \left( U_n \to \frac{\zeta+\zeta_n}{\zeta-\zeta_n} U_n \right) ,
\nonumber \\
{\rm det} (\omega+D) & = & {\rm det} (\omega+B) \left( U_n \to   - \frac{\zeta+\zeta_n}{\zeta-\zeta_n} U_n \right) .
\label{28cc}
\end{eqnarray}
Now  the potential can be chosen to be real, $\Delta=S$, and  these solutions describe the scattering of real kinks with 
twist angle $\pi$. Furthermore, they are in fact characterized by solutions to the Sinh-Gordon model, as $S(x, t)$ satisfies the nonlinear 
equation \cite{13b}
\begin{equation}
\partial \bar{\partial} ( \ln S^2 ) =
\frac{1}{2} \left( S^2-\frac{1}{S^2} \right)  =  \sinh ( \ln S^2 ) .
\label{28d}
\end{equation}

\subsection{Master equation}
\label{sect3d}
Our general transparent solutions do not satisfy the Sinh-Gordon equation (\ref{28d}). The most general
nonlinear partial differential equation satisfied by the transparent potential 
$\Delta$ relates it to the common denominator, ${\rm det} (\omega + B)$, of the three expressions in (\ref{28}):
\begin{equation}
4 \partial \bar{\partial} \ln {\rm det}(\omega + B) = 1 - |\Delta|^2.
\label{29}
\end{equation} 
We refer to this equation as ``master equation", since in large $N$ fermionic field theories $\Delta$ plays a role similar to  
the ``master field" in large $N$ gauge theories \cite{14}. To prove (\ref{29}), we evaluate its left hand side, 
using $\ln {\rm det} = {\rm Tr}\, \ln$,
\begin{equation}
4 \partial \bar{\partial}\, {\rm ln}\, {\rm det} (\omega+B) = 4 {\rm Tr} \left(  \frac{1}{\omega+ B} \partial \bar{\partial} B
  - \frac{1}{\omega+B} \partial  B \frac{1}{\omega + B} \bar{\partial}B \right).
\label{30}
\end{equation}
With the help of (\ref{18}) and the 2nd derivative
 \begin{equation}
\partial \bar{\partial} B = \frac{i}{4} \left( e f^{\dagger} - f e^{\dagger} \right),
\label{31}
\end{equation}
we find
\begin{eqnarray}
4 \partial \bar{\partial} \ln {\rm det}(\omega + B) & = & i f^{\dagger}\frac{1}{\omega + B} e - i e^{\dagger} \frac{1}{\omega + B} f 
 -  e^{\dagger}  \frac{1}{\omega+B} f f^{\dagger}  \frac{1}{\omega+B} e
\nonumber \\
& =  &  1 - |\Delta|^2,
\label{32}
\end{eqnarray}
thus confirming Eq.~(\ref{29}). The master equation may be viewed as the generalization of the following equation of KM \cite{1},
\begin{eqnarray}
V(x) & = &  -  \partial_x^2  \ln  {\rm det} (1+ A),
\nonumber \\
A_{ij} & = & \sqrt{a_i a_j} \frac{e^{(\kappa_i+ \kappa_j)x}}{\kappa_i+\kappa_j}.
\label{33}
\end{eqnarray}
Restricting  by the algebraic conditions of diagonal $\omega$ and spectral parameters lying on the unit circle,
and expanding $|\Delta| \approx 1+V$ in a non-relativistic limit,  the master equation (\ref{29}) reduces to the log det form of the
Kay-Moses transparent potential in (\ref{33}).
\subsection{Spatial asymptotics of potential and continuum spinors}\label{sect3e}
The determinantal forms of $\Delta, \chi_1, \chi_2$ derived in the subsections \ref{sect3b} and \ref{sect3c} are particularly convenient for 
extracting the asymptotics of the potential and associated spinors for $x \to \pm \infty$. With each bound state pole $\zeta_n$, 
we can associate a complex momentum
\begin{equation}k_n = \frac{1}{2} \left( \zeta_n - \frac{1}{\zeta_n} \right).
\label{34}
\end{equation}
Let us assume that ${\rm Im}\, k_n >0$ for all $n$. Then the matrix $B$ blows up exponentially for $x\to \infty$, and vanishes 
for $x\to - \infty$. Thus the matrices $A,B,C,D$ dominate over the matrix $\omega$ for $x\to \infty$, whereas $\omega$ dominates 
over $A,B,C,D$ for $x\to -\infty$. The potential $\Delta$ satisfies
\begin{equation}\lim_{x\to - \infty}\Delta =  1 , \quad \lim_{x \to \infty}\Delta  = \frac{{\rm det} A}{{\rm det} B} =
 \prod_{n=1}^N \frac{\zeta_n}{\zeta_n^*}.
\label{35}
\end{equation}
This shows that $\Delta$ has a chiral twist. It starts out at the point ($S=1,P=0$) on the chiral (unit) circle at $x\to - \infty$, and 
reaches another point on the chiral circle at $x \to +\infty$, say $e^{i\Theta}$. According to (\ref{35}), the chiral twist angle $\Theta$ 
can be computed by simply adding up the phases of all bound state pole parameters $\zeta_n$,
\begin{equation}
\Theta = 2 \sum_{n=1}^N \theta_n, \qquad \zeta_n = |\zeta_n| e^{i\theta_n}.
\label{36}
\end{equation}
The spinor components have the asymptotic behavior
\begin{eqnarray}
\lim_{x\to - \infty}\chi_1 & = &   1 , \quad \lim_{x \to \infty}\chi_1  = \frac{{\rm det} C}{{\rm det} B}
 = \prod_{n=1}^N \frac{\zeta-\zeta_n^*}{\zeta-\zeta_n},
\label{37} \\
\lim_{x\to - \infty}\chi_2 & = &   1 , \quad \lim_{x \to \infty}\chi_2  = \frac{{\rm det} D}{{\rm det} B} =\prod_{n=1}^N \frac{\zeta_n}{\zeta_n^*}
  \frac{\zeta-\zeta_n^*}{\zeta-\zeta_n}.
\label{38}
\end{eqnarray} 
The product in (\ref{37}) can be identified with the fully factorized, unitary transmission amplitude $T(\zeta)$ with the expected
 pole structure,
\begin{equation}
T(\zeta) =  \prod_{n=1}^N \frac{\zeta-\zeta_n^*}{\zeta-\zeta_n}, \quad |T(\zeta)|=1.
\label{39}
\end{equation}
The extra factors in the product in (\ref{38}) are necessary to account for the chiral twist of the potential $\Delta$ at $x \to \infty$, which 
also affects the spinors. 
\subsection{Normalization and orthogonality of bound states}
\label{sect3f}
According to Eqs.~(\ref{15}, \ref{16}), the spinor components of the $n$-th bound state spinor $\phi_n$ are given by
\begin{equation}
\phi_{1,n} = \left( \frac{1}{\omega + B} e \right)_n, \quad  \phi_{2,n} = - \left( \frac{1}{\omega + B} f \right)_n.
\label{40}
\end{equation}
Consider the ``density matrix"  
\begin{eqnarray}
\phi_n^{\dagger} \phi_m & = &  \phi_{1,n}^* \phi_{1,m} +  \phi_{2,n}^* \phi_{2,m}
\nonumber \\
& = & \left( \frac{1}{\omega+B} \left( e e^{\dagger} + f f^{\dagger} \right) \frac{1}{\omega+B} \right)_{mn} .
\label{41}
\end{eqnarray}
Here, the word ``density matrix" refers to the bound state labels $n,m$, with the understanding that all spinors are evaluated 
at the same space-time arguments. Since, accoording to (\ref{18}),
\begin{equation}
\partial_x B = (\partial + \bar{\partial})B = \frac{1}{2} \left( e e^{\dagger}+ f f^{\dagger} \right),
\label{42}
\end{equation}
the density matrix can be written as a total spatial derivative,  
\begin{equation} 
\phi_n^{\dagger} \phi_m = - 2 \partial_x \left( \frac{1}{\omega + B} \right)_{mn}. 
\label{43} 
\end{equation}
This is very useful, as it enables us to evaluate the normalization matrix trivially,
\begin{equation}
R_{nm} = \int_{-\infty}^{\infty} dx \, \phi_n^{\dagger} \phi_m = - 2 \left. \left(\frac{1}{\omega + B}\right)_{mn} \right|_{x=-\infty}^{x=+\infty} = 
2 \left( \omega^{-1} \right)_{mn}.
\label{44}
\end{equation}
In the last step, we have once again assumed that the bound state poles have ${\rm Im}\, k_n>0$. Orthonormalized bound 
states $\hat{\phi}_n$ can be constructed from the $\phi_n$ via
\begin{equation}
\hat{\phi}_n = \sum_{m} C_{nm} \phi_m, \quad \int dx \,\hat{\phi}_n^{\dagger} \hat{\phi}_m = \delta_{n,m}.
\label{45}
\end{equation}
The following condition for the matrix $C$ follows from Eqs.~(\ref{44}, \ref{45}),
\begin{equation} 2 C \omega^{-1} C^{\dagger} = 1.
\label{46}
\end{equation}
This condition is needed for applying the present results to integrable quantum field theories in the large $N$ limit \cite{12,13}.
\section{Summary and conclusions}
\label{sect4}
We have found a simple algebraic characterization of a large class of transparent, scalar-pseudoscalar Dirac potentials in one
dimension. The novel feature is the fact that these potentials are time-dependent. Whether our results exhaust all transparent Dirac 
potentials remains to be seen. They contain all previously found transparent potentials,  as special cases, be they static Dirac potentials, 
or static and non-static Schr\"odinger potentials, and also provide new classes of time-dependent Dirac potentials associated with multi-baryons and multi-breathers
 \cite{12,13}. Our expressions in terms of determinants are natural generalizations of the original Kay-Moses results for static 
Schr\"odinger potentials. In view of the tremendous generalization from the static Schr\"odinger potentials of KM more 
than half a century ago, it is very surprising that the derivation in the time-dependent Dirac case is hardly more 
complicated. In fact, we were strongly guided by the KM work when setting up the proof for the time-dependent, relativistic case. 

Our results open the door for solving a variety of dynamical problems in large $N$ fermionic quantum field theories. These include 
scattering problems of arbitrary number of baryonic bound states, but also breathers of any degree of complexity and their interactions.
According to general experience, the self-consistent potentials in Hartree-Fock or TDHF approaches to integrable models  are always
transparent, so that the present study gives a good starting point for identifying self-consistent solutions \cite{12,13}. However, 
transparent potentials may also have other applications than the one to integrable quantum field theories, such as for example in 
condensed matter physics, quantum optics, or cold atom physics. 
\section*{Acknowledgement}
G.D. acknowledges support from DOE grants DE-FG02-92ER40716 and DE-FG02-13ER41989, and the ARC Centre of Excellence in
Particle Physics at the Terascale and CSSM, School of Chemistry and Physics, University of Adelaide. 
M.T. thanks Oliver Schnetz for stimulating discussions and the DFG for financial support under grant TH 842/1-1.  


\begin{thebibliography}{99}
\bibitem{1}I. Kay and H. E. Moses, ``Reflectionless transmission through dielectrics and scattering potentials'', J. Appl. Phys. {\bf 27}, 1503 (1956).
\bibitem{1a}K. Chadan and P. C. Sabatier, {\it Inverse Problems in Quantum Scattering Theory}, 2nd Ed. (Springer, Berlin, 1989).
\bibitem{2}Y. Nogami and C. S. Warke, ``Soliton solutions of multicomponent nonlinear Schr\"odinger equation'', Phys. Lett. A {\bf 59}, 251 (1976).
\bibitem{3}F. M. Toyama, Y. Nogami, Z. Zhao, ``Relativistic extension of the Kay-Moses method for constructing transparent potentials in
 quantum mechanics'', Phys. Rev. A {\bf 47}, 897 (1993).
\bibitem{4}D. J. Gross and A. Neveu,   ``Dynamical Symmetry Breaking in Asymptotically Free Field Theories,''  Phys. Rev. D {\bf 10}, 3235 (1974).
\bibitem{5}R. F. Dashen, B. Hasslacher and A. Neveu,  ``Semiclassical Bound States in an Asymptotically Free Theory,'' 
Phys. Rev. D {\bf 12}, 2443 (1975).
\bibitem{6}J. Feinberg, ``All about the static fermion bags in the Gross-Neveu model,'' Ann. Phys. (N.Y.) {\bf 309}, 166 (2004),
\hhref{hep-th/0305240}.  
\bibitem{7}Y. Nogami and F. M. Toyama, ``Reflectionless potential for the one-dimensional Dirac equation: pseudoscalar potentials'',
Phys. Rev. A {\bf 57}, 93 (1998).
\bibitem{8}D. A. Takahashi, S. Tsuchiya, R. Yoshii, M. Nitta, ``Fermionic solutions of chiral Gross-Neveu and Bogoliubov-de Gennes 
systems in nonlinear Schr\"odinger hierarchy,''  Phys.\ Lett.\ B {\bf 718}, 632 (2012), \hhref{1205.3299 [cond-mat.supr-con]}.  
\bibitem{9}D. A. Takahashi and M. Nitta,  ``Self-consistent multiple complex-kink solutions in Bogoliubov-de Gennes and chiral Gross-Neveu
 systems,'' Phys.\ Rev.\ Lett.\  {\bf 110}, 131601 (2013), \hhref{1209.6206 [cond-mat.supr-con]}.  
\bibitem{10}D. A. Takahashi and M. Nitta, ``On reflectionless nature of self-consistent multi-soliton solutions in Bogoliubov-de Gennes and 
chiral Gross-Neveu models,'' \hhref{1307.3897 [cond-mat.supr-con]}.  
\bibitem{11}S.-S. Shei,  ``Semiclassical Bound States in a Model with Chiral Symmetry,'' Phys. Rev. D {\bf 14}, 
535 (1976).
\bibitem{12}G.~V.~Dunne and M.~Thies,  ``Time-Dependent Hartree-Fock Solution of Gross-Neveu models: Twisted Kink 
Constituents of Baryons and Breathers,'' \hhref{1306.4007 [hep-th]}.  
\bibitem{13}G. V. Dunne and M. Thies, ``Full Time-Dependent Hartree-Fock Solution of Large N Gross-Neveu models'',
to be published.
\bibitem{13a}
K. B. Petersen and M. S. Pedersen, {\it The Matrix Cookbook}, Version November 15, 2012, p. 6.
\bibitem{13b}
C. Fitzner and M. Thies, ``Exact solution of N baryon problem in the Gross-Neveu model", Phys. Rev. D {\bf 83}, 085001 (2011),
\hhref{1010.5322 [hep-th]}.
\bibitem{14}E. Witten, in {\em Recent developments in gauge theories}, 
1979 Cargese lectures, ed. G. 't~Hooft et al., Plenum Press, N.Y. (1980).
\end{thebibliography}
\end{document}